\documentclass[10pt,twocolumn,letterpaper]{article}

\usepackage{iccv}              

\definecolor{iccvblue}{rgb}{0.21,0.49,0.74}
\usepackage[pagebackref,breaklinks,colorlinks,allcolors=iccvblue]{hyperref}
\usepackage{subcaption}
\usepackage{graphicx}
\def\paperID{*****} 
\def\confName{ICCV}
\def\confYear{2025}

\title{Evaluating and Improving the Effectiveness of Synthetic Chest X-Rays for Medical Image Analysis}

\author{
Eva Prakash\textsuperscript{1}
\and
Jeya Maria Jose Valanarasu\textsuperscript{1}
\and
Zhihong Chen\textsuperscript{1}
\and
Eduardo Pontes Reis\textsuperscript{1}
\and
Andrew Johnston\textsuperscript{1}
\and
Anuj Pareek\textsuperscript{1}
\and
Christian Bluethgen\textsuperscript{1}
\and
Sergios Gatidis\textsuperscript{1}
\and
Cameron Olsen\textsuperscript{1}
\and
Akshay Chaudhari\textsuperscript{1}
\and
Andrew Ng\textsuperscript{1}
\and
Curtis Langlotz\textsuperscript{1}\\
\textsuperscript{1}Stanford University, Stanford, CA, USA\\
}

\begin{document}
\maketitle
\begin{abstract}
In this work, we explore best-practice approaches for generating synthetic chest X-ray images and augmenting medical imaging datasets to optimize the performance of deep learning models in downstream tasks like classification and segmentation. We utilize a latent diffusion model to condition the generation of synthetic chest X-rays on text prompts and/or segmentation masks. We explore methods such as using a proxy model and incorporating radiologist feedback to improve the quality of synthetic data. These synthetic images are generated from relevant disease information or geometrically-transformed segmentation masks and added to ground truth training set images from the CheXpert \cite{irvin2019chexpert}, CANDID-PTX\cite{feng2021curation}, SIIM \cite{siim}, and RSNA Pneumonia \cite{stein2018rsna} in order to measure improvements in classification and segmentation model performance on the test sets. F1 and Dice scores are used to evaluate classification and segmentation, respectively. Across all experiments, the synthetic data we generate results in a maximum mean classification F1 score improvement of 0.15 (CI: 0.10, 0.20; P=0.0031) compared to using only real data. For segmentation, the maximum Dice score improvement is 0.14 (CI: 0.11, 0.18; P=0.0064). We find that best practices for generating synthetic chest X-ray images for downstream tasks include conditioning on single-disease labels or geometrically-transformed segmentation masks, as well as potentially using proxy modeling to fine-tune such generations.
\end{abstract}
\section{Introduction}
\label{sec:intro}
\begin{table*}[t]
  \centering
  \begin{tabular}{@{}lll@{}}
    \toprule
    \textbf{Dataset} & \textbf{Train/Test Size} & \textbf{Conditions} \\
    \midrule
    CheXpert \cite{irvin2019chexpert} & 191,229 / 235 & Atelectasis, cardiomegaly, consolidation, edema, pleural effusion \\
    CANDID-PTX \cite{feng2021curation} & 17,237 / 2,000 & Lungs, heart, pneumothorax, rib fractures, chest tubes \\
    SIIM \cite{siim} & 1,903 / 476 & Pneumothorax \\
    RSNA Pneumonia \cite{stein2018rsna} & 21,885 / 4,801 & Pneumonia \\
    \bottomrule
  \end{tabular}
  \caption{Overview of datasets used for training and evaluation. CANDID-PTX and SIIM were used for pneumothorax segmentation, CheXpert for multi-label classification, and RSNA Pneumonia for binary classification.}
  \label{tab:datasets}
\end{table*}
Developing AI tools for medical image analysis is inherently challenging due to limited data availability and the need for carefully labeled medical datasets. The scarcity of publicly available medical imaging datasets arises from constraints such as logistical challenges, patient privacy requirements, and the high cost of expert annotations. Synthetic data generation has emerged as a promising approach to address these limitations and can be used to augment datasets for disease segmentation and classification tasks. While previous studies have demonstrated the feasibility of generating synthetic data for chest X-rays using earlier methods like generative adversarial networks (GANs) \cite{goodfellow2014gan}, these models often struggle to produce clinically meaningful details, which limits their effectiveness in downstream applications \cite{frid2018gan, shin2018medical}. Recent diffusion-based approaches, such as vision-language foundation models for chest X-ray generation, have since advanced synthetic image quality by producing more realistic, high-resolution outputs \cite{bluethgen2024vision}. However, these diffusion models often lack the ability to generate paired images and labels, limiting their utility for segmentation tasks where such pairs are essential \cite{kazerouni2023diffusion}. Additionally, the effects of the quality and precision of synthetic medical images on downstream tasks, such as classification and segmentation, remain under-explored.
\newline To address these challenges, this study uses a latent diffusion model \cite{rombach2022ldm} based on the ControlNet architecture \cite{controlnet} designed to generate synthetic chest X-rays conditioned on both text prompts and segmentation masks. Unlike prior models, this framework produces paired images and labels, enhancing its applicability for a broader range of downstream tasks, including segmentation and classification. We examined several hypotheses about synthetic data generation for medical imaging. We hypothesize that synthetic data can enhance performance in data-scarce environments, especially when the synthetic data increases the volume of the dataset. We also explore whether producing visually realistic, low-hallucination images leads to improved outcomes in classification and segmentation tasks. Finally, we investigate the impact of selecting synthetic images based on a pseudo-model mimicking the downstream task, positing that this approach can improve synthetic data selection and optimize performance. 
\newline Through these contributions, this work aims to deepen the understanding of synthetic data generation in medical imaging and establish effective strategies for leveraging synthetic images in classification and segmentation tasks under data-limited conditions.

\section{Related Work}
\subsection{Medical Image Generation and Synthetic Data Augmentation}
In the medical domain, synthetic data is critical in enhancing the quality of limited image datasets. Derm-T2IM uses synthetic skin lesion data generated with a Stable Diffusion model to improve the performance and robustness of real-world skin lesion machine learning models \cite{dermt2im}. In a similar vein, Khader et al. use self-supervised pre-training on diffusion models to synthesize high-quality MRI and CT scans and enhance breast segmentation model performance with limited data \cite{khader}. However, unlike our model, these models are not conditional. Bluethgen et al. adapts a pre-trained latent diffusion model to generate synthetic chest X-ray images conditioned on summarized radiology reports, leading to improvements in downstream classifier performance when such images are utilized for data augmentation \cite{bluethgen2024vision}. However, this model does not include mask conditioning. In addition, none of these models are finetuned with radiologist preference scores. Our model aims to explore some of these shortcomings.

\subsection{Diffusion Models and Instruction Tuning}
Diffusion models \cite{ddpm, rombach2022ldm} have enabled controllable and high-fidelity image generation and are increasingly popular in medical imaging \cite{khader, dermt2im}. Extensions such as ControlNet \cite{controlnet} add conditional controls such as edge maps or semantic segmentations to generated images. We build upon pretrained latent diffusion models (LDMs) as they allow scalable and efficient fine-tuning.

Human preferences have been used to improve the outputs of pretrained generative models by giving sets of samples to human annotators to rank based on quality; the rankings are then integrated into the loss functions of generative models so that their outputs can be finetuned in the direction of the provided rankings. Reinforcement learning from human feedback (RLHF) \cite{rlhf} involves training a separate reward model that learns to assign numerical quality scores to data based on learned patterns from human rankings and using the trained reward model to serve as a regularizer when finetuning generative large language models. However, RLHF has the drawbacks of having sensitivity to hyperparameters as well as the inefficiency of training a separate reward model. Methods such as direct preference optimization (DPO) \cite{rafailov2023direct} and preference ranking optimization \cite{prefranking} build upon these shortcomings by directly utilizing human rankings in the loss functions when finetuning large language models, although since there is no explicit reward model distribution learned, these methods run the risk of overfitting.

\begin{figure*}[t]
    \centering
    \includegraphics[width=0.6\textwidth]{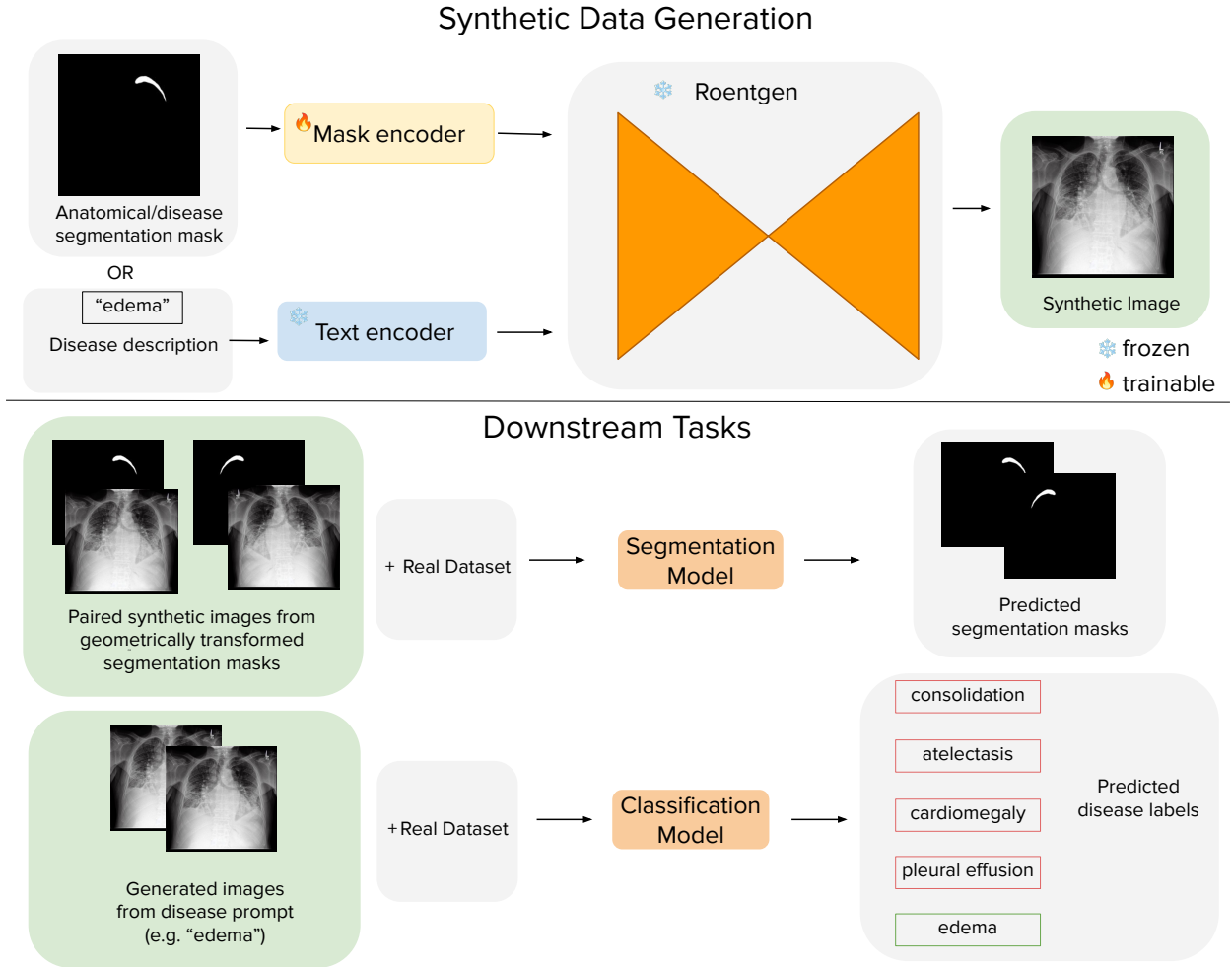}
    \caption{Pipeline for synthetic data generation and downstream classification. We generate synthetic images using segmentation masks or text prompts inputted into the ControlNet architecture pre-initialized with the Bluethgen et al. model. These synthetic images are then used for augmenting existing training sets used for segmentation or classification tasks to predict anatomical masks or disease labels.}
    \label{fig:main}
\end{figure*}
\section{Methods}
\label{sec:formatting}

\subsection{Synthetic chest X-rays can be generated by conditioning on custom mask and reports to mimic diseases and conditions}

Bluethgen et al. \cite{bluethgen2024vision} adopt a pre-trained latent diffusion architecture to generate synthetic chest X-ray images conditioned on summarized radiology reports. In this work, we show that a latent diffusion model can be easily extended to generate synthetic chest X-rays conditioned on more than just text. To do this, our latent diffusion model framework utilizes the ControlNet architecture \cite{controlnet}, which integrates visual conditioning (e.g. segmentation masks) into pretrained, frozen text-to-image latent diffusion models like Stable Diffusion \cite{rombach2022ldm}. ControlNet integrates these conditions by cloning the frozen parameters to create trainable counterparts that accept conditioning vectors. These trainable counterparts are linked through zero-initialized convolution layers, enabling the model to learn new conditional patterns without affecting its original generative abilities. We reuse the model weights of Bluethgen et al., as it already is trained on a large-scale dataset of chest X-rays and reports.  With this, we create a flexible framework as seen in Figure \ref{fig:main}, which can take in both a mask and a text prompt to create accurate synthetic chest X-rays grounded by the input conditions we provide.

\subsection{Datasets}

 Details regarding all datasets used in our experiments can be found in Table \ref{tab:datasets}. We used the CANDID-PTX chest X-ray dataset for training our diffusion model. A total of 17,237 randomly selected images were used for training, while the remaining 2,000 images were reserved for evaluation. These 2,000 held-out images were then employed for downstream pneumothorax segmentation, with the data split randomly into an 80/20 ratio for training and testing. Following the MGCA \cite{wang2022multi} paper, we utilized the validation set of CheXpert as our test set. To more directly test segmentation performance, we only use the SIIM images with nonzero segmentation masks and randomly split these images 80/20 for training and testing. Following the MGCA \cite{wang2022multi} paper, we also randomly split the RSNA Pneumonia training set 80/20, given the public availability of only the training annotations.

\subsection{Model Architecture and Training}

Our latent diffusion model framework incorporates segmentation mask visual conditioning into the pre-initialized Bluethgen et al. model \cite{bluethgen2024vision} and its corresponding text encoder using the ControlNet framework.  While training our model on segmentation masks, text conditions were set to ``N/A,'' prompting the model to convert text conditions to zero vectors. During sampling, if ``N/A'' was provided as a text condition or an all-black segmentation mask was used, the corresponding condition vectors were zeroed. This approach enabled the trained ControlNet model to generate synthetic data from any combination of text and mask conditions.

We train our model with the ControlNet loss function. It takes in an input image $z_0$ and progressively adds random noise to produce a noisy version $z_t$, where $t$ is the given time step. Given the mask condition $c_f$, text condition $c_t$, and time step $t$, the model trains by learning a noise prediction network $\epsilon_\theta$ to predict the noise $\epsilon$ added to $z_t$, as shown in Equation 1:

\begin{equation}
L = \mathbb{E}_{z_0,t,c_t,c_f,\epsilon \sim \mathcal{N}(0,1)} \left[
\left\| \epsilon - \epsilon_\theta(z_t, t, c_t, c_f) \right\|_2^2
\right],
\label{eq:controlnet}
\end{equation}

We trained the model on eight NVIDIA A4000 GPUs. Gradient accumulation was set to 16, and the learning rate was $1 \times 10^{-4}$. Other packages and hyperparameters remained consistent with those in the original ControlNet implementation.

\subsection{Improving the synthetic data}

\subsubsection{Proxy Model}

 As shown in Figure \ref{fig:proxymodel}, as a proxy model, we employ BiomedCLIP \cite{eswaran2022biomedclip} to identify and filter medical images depicting one of the five diseases outlined in the CheXpert classification tasks (i.e. "cardiomegaly", "pleural effusion," "edema," "atelectasis", and "consolidation"). Our approach calculates cosine similarity scores between features extracted from each image and the corresponding disease text label. Images from a specified dataset are processed individually, and similarity scores are computed to assess how closely each image aligns with the target label. To ensure specificity, we apply a 90th-percentile threshold, selecting images with the highest similarity scores. This percentile-based filtering metric is instrumental in isolating images with the strongest relevance to the specified medical condition. The model was trained for an additional 9,000 steps on the filtered images.

\subsubsection{Radiologist Feedback}
\begin{figure*}[t]
    \centering
    \includegraphics[width=0.7\textwidth]{proxy_xl.pdf}
    \caption{Proxy model-based refinement for synthetic data. We select high-quality synthetic images through cosine similarity between the disease prompt and image using BioMedCLIP and fine-tune our diffusion model on the high-quality images. Using the 1\% setting of the CheXpert dataset, given the large size of the dataset, we find that fine-tuning improves CheXpert classification performance the most, and simply using the filtered high-quality images for downstream classification improves performance the least.}
    \label{fig:proxymodel}
\end{figure*}

For reinforcement learning fine-tuning, as shown in Figure \ref{fig:dpo}, we collected preferences from radiologists to enhance the model's outputs. We assembled approximately 200 sets of four sampled images each, with the distribution of condition types as follows: 75\% text-only conditioning (T2I), 15\% mask-only conditioning (M2I), and 10\% combined text and mask conditioning (TM2I). The percentages were chosen based on the sampling diversity allowed by each condition type.

Text conditions for validation were generated from the impressions section of the MIMIC-CXR validation split and the held-out CANDID-PTX reports. These were summarized into concise five-word captions using GPT-4 with a standardized prompt to ensure consistency and avoid bias:

\begin{quote}
\emph{``Based on these chest X-ray reports, please write a five-word caption with the main finding. Don't make comparisons with previous studies, so do not use words such as 'unchanged', 'improved', 'worsened', 'no change', 'increased', 'decreased', etc. in the caption. Don't use commas or quotation marks in the caption. If it is normal or no problems are detected, just return 'Normal' as the caption.''}
\end{quote}

The image sets were hosted on Gradio, and four radiologists independently scored each image set on a scale from 0 (low quality) to 5 (high quality). For images with combined conditions, separate scores were collected for text and mask condition quality. The data were anonymized, and the radiologists were blinded to the source of the images.

We employed Direct Preference Optimization (DPO) \cite{rafailov2023direct} to utilize these preferences in model fine tuning. DPO operates directly on the preference scores, and the loss function is defined as:

\begin{equation}
\begin{split}
L_{\text{DPO}}(\pi_\theta; \pi_{\text{ref}}) 
= -\mathbb{E}_{(c, y_w, y_l) \sim D} \Big[
\log \sigma \Big(
\beta \log \frac{\pi_\theta(y_w \mid c)}{\pi_{\text{ref}}(y_w \mid c)} \\
\qquad\qquad\qquad
- \beta \log \frac{\pi_\theta(y_l \mid c)}{\pi_{\text{ref}}(y_l \mid c)}
\Big) \Big]
\end{split}
\end{equation}

where $\beta$ is a scaling factor, $\pi_\theta$ is the model policy parameterized by $\theta$, $\pi_{\text{ref}}$ is the reference model policy, $(c,y_w,y_l)$ represents an example in dataset $D$ where $y_w$ is a more preferred sample, $y_l$ is a less preferred sample, and $c$ is the corresponding condition. Separate losses are calculated for each condition type (text or mask) and zeroed or averaged based on the presence or absence of each condition. We use the parameters of the DPO paper, except for a learning rate of $1e^{-6}$ and a gradient accumulation of 1. The model was trained for an additional 4,000 steps on the collected preference data.

\subsection{Downstream Evaluation}

To assess the effectiveness of synthetic data in enhancing performance on out-of-distribution datasets, we conducted a series of downstream segmentation and classification experiments across multiple disease conditions. For segmentation, we utilized the UNet++ \cite{unet++} architecture, an advanced encoder-decoder network with nested dense skip pathways, to segment pneumothorax regions. We augmented the CANDID-PTX and SIIM training datasets by applying geometric transformations—including dilation, erosion, horizontal flipping, and translation—to ground truth masks. These modified masks were then inputted into the model to generate new synthetic data paired with each mask, thus expanding the datasets with diversified synthetic instances.

In the classification domain, multi-class disease classification was performed using the MGCA \cite{wang2022multi} architecture to identify pleural effusion, atelectasis, edema, cardiomegaly, and consolidation on the CheXpert dataset. Single-class classification for pneumonia detection was carried out using a ResNet-50 \cite{he2015deep} model. The training datasets for both tasks were augmented with synthetic images, with each synthetic instance corresponding to a single disease label to enhance model specificity. Model training parameters followed those outlined in the respective original architecture publications, ensuring consistency in evaluation.

We used F1 scores to evaluate classification and Dice scores for segmentation. This study posits three hypotheses: (1) synthetic data augmentation enhances performance in both segmentation and classification tasks, particularly in data-scarce settings; (2) increasing the volume of synthetic data correlates positively with model performance; (3) the generation of visually realistic, low-hallucination synthetic images, which can be generated using feedback from either experts or proxy models, somewhat contributes to improved task outcomes.

\subsection{Statistical Analysis}
One-tailed t-tests with Bonferroni correction of the performance differences between ground truth and synthetic-data-augmented downstream models were used to assess the statistical significance of performance improvements with synthetic data. We used two trials to calculate the mean values for each experimental setting and used a significance level of $P < 0.05$ to determine general improvements with synthetic data.

\begin{figure*}[t]
    \centering
    \includegraphics[width=0.7\textwidth]{dpo_xl.pdf}
    \caption{Radiologist feedback integration for synthetic image generation. We collect synthetic images from our diffusion model pipeline and solicit radiologist feedback by ranking the quality of sets of images. We finetune our model on the rankings to refine the model in the direction of expert preferences. Using the 10\% setting of the RSNA Pneumonia dataset, we found that while finetuned synthetic images improve pneumonia classification over real images, our original synthetic data still performs better.}
    \label{fig:dpo}
\end{figure*}
\section{Experiments}
\begin{figure*}[t]
    \centering
    \begin{subfigure}[t]{0.7\textwidth}
        \centering
        \includegraphics[width=\linewidth]{candid_p_results.pdf}
        \caption{CANDID-PTX: Segmentation}
    \end{subfigure}
    \hfill
    \begin{subfigure}[t]{0.7\textwidth}
        \centering
        \includegraphics[width=\linewidth]{siim_p_results.pdf}
        \caption{SIIM: Segmentation}
    \end{subfigure}

    \vspace{1em}

    \begin{subfigure}[t]{0.7\textwidth}
        \centering
        \includegraphics[width=\linewidth]{rsna_p_results.pdf}
        \caption{RSNA: Classification}
    \end{subfigure}
    \hfill
    \begin{subfigure}[t]{0.7\textwidth}
        \centering
        \includegraphics[width=\linewidth]{chexpert_p_results.pdf}
        \caption{CheXpert: Classification}
    \end{subfigure}
    \caption{Downstream segmentation and classification performance with real and synthetic data. We find that the addition of synthetic data improves segmentation (CANDID-PTX, SIIM) and classification (RSNA, CheXpert) performance, particularly with limited ground truth data.}
    \label{fig:results}
\end{figure*}
\subsection{Synthetic chest X-rays help improve model performance on classification and segmentation tasks}

We tested whether synthetic data can improve the performance of both binary and multilabel disease classification tasks. For binary pneumonia classification, we trained a ResNet 50 \cite{he2015deep} model to predict the presence or absence of pneumonia in the RSNA Pneumonia dataset. For multilabel classification, we trained the MGCA model \cite{wang2022multi} on the CheXpert dataset to classify five disease labels following the MGCA \cite{wang2022multi} paper: atelectasis, edema, consolidation, cardiomegaly, and pleural effusion. We first evaluated the F1 scores of both on the real test sets. Then synthetic data was added to augment the training datasets at 2x, 5x, 10x, and 25x the size of the original real training data. Finally we re-evaluated the trained models on the same real test sets.

Binary classification for pneumonia showed substantial improvements with the inclusion of synthetic data (Figure \ref{fig:results}, Plot 3). Synthetic pneumonia chest X-rays were generated using a single-disease prompt (i.e. "pneumonia"). At 1\% ground truth data availability, the mean F1 score improved from 0.0104 (real-only training) to 0.1541 with 25x synthetic augmentation. At 10\% ground truth data, the F1 score increased from 0.0950 to 0.2760 with 10x synthetic data. The benefits of synthetic data diminished at 100\% ground truth data, where the F1 score plateaued (0.326 with real data vs. 0.344 with 2x synthetic data).

In multilabel classification on CheXpert dataset, we also found it was most effective to generate the synthetic data using single-disease prompts (e.g. "cardiomegaly"). The synthetic data was used to balance all classes to the count of the most frequent disease class. At 1\% ground truth data, the mean F1 score increased from 0.209 with real data to 0.289 with 2x synthetic data. At 10\% ground truth data, the F1 score improved from 0.255 to 0.323 with 2x synthetic data. At 100\% ground truth data, synthetic data provided marginal improvements, with the F1 score increasing from 0.310 to 0.354 with balanced 2x synthetic augmentation (Figure \ref{fig:results}, Plot 4). These results demonstrate that the addition of synthetic data can address class imbalance and significantly improve multilabel classification performance, especially in low-resource scenarios.


For segmentation, the UNet++ \cite{unet++} architecture was employed to segment pneumothorax regions in the CANDID-PTX and SIIM datasets. Geometric transformations, including dilation, erosion, horizontal flipping, and translation, were applied to ground truth masks to generate diverse variations. These augmented masks served as input conditions to the diffusion model, which produced paired synthetic images and labels. Performance was evaluated on the real test sets, with synthetic data augmenting the training sets at 2x, 5x, 10x, and 25x the size of the real training data.


At 1\% ground truth data availability, the mean Dice score improved from 0.0484 with real data to 0.0601 with 25x synthetic augmentation. At 10\% ground truth data, synthetic data improved the Dice score from 0.0569 to 0.196 with 10× synthetic data. At 100\% ground truth data, segmentation performance gains were marginal, with Dice scores increasing from 0.320 to 0.334 with 2x synthetic data (Figure \ref{fig:results}, Plot 1).

Similar trends were observed for the SIIM dataset. At 1\% ground truth data, the Dice score increased from 0.2072 with real data to 0.2550 with 25x synthetic augmentation. At 10\% ground truth data, the Dice score improved from 0.237 to 0.251 with 10x synthetic data. At 100\% ground truth data, the improvements were minimal, with Dice scores increasing from 0.374 to 0.382 with 2x synthetic data (Figure \ref{fig:results}, Plot 2). These findings show that synthetic data effectively improves segmentation performance, especially in data-scarce settings. However, the benefit diminishes as the availability of real data increases.

Overall, the addition of synthetic data improves the mean Dice score for segmentation tasks and mean F1 score for classification tasks, as demonstrated by the performance increase across almost all conditions when comparing the use of only ground truth data versus combinations of ground truth and synthetic data. This suggests that synthetic data can effectively augment existing datasets, particularly in settings where limited ground truth data is available.

\subsection{More synthetic data  results in better model performance}
Increasing the proportion of synthetic data further boosts downstream segmentation and classification performance, particularly in cases with minimal ground truth data (e.g., 1\% and 10\% ground truth data). Models trained with progressively more synthetic data tend to achieve higher mean Dice and F1 scores compared to models trained with lower levels of synthetic augmentation. This progressive improvement underscores the value of synthetic data in augmenting small datasets, where limited real data may be insufficient for achieving optimal performance. However, this trend plateaus at higher data availability (e.g., 100\% ground truth data), indicating that synthetic data is most beneficial when ground truth data is scarce and may offer diminishing returns when real data availability is robust.

\subsection{Enhancing synthetic data with a proxy model boosts performance}
Chest X-rays exhibit fine-grained nuances and complexities that are challenging to capture in synthetic data, making fine-tuning helpful for improving representations. Radiologist feedback helps steer the model toward established medical knowledge, offering valuable guidance to refine the generation process. However, collecting such expert input at scale remains difficult. As a result, medical foundation models can serve as proxies for radiologists by efficiently offering metrics like cosine similarity between captions and images to evaluate the quality of synthetic generations, providing a scalable alternative to direct involvement from clinical experts.

We employed a proxy model approach using BioMedCLIP \cite{eswaran2022biomedclip} to filter high-quality synthetic images based on cosine similarity between disease prompts and generated images. The filtered synthetic data was then used to fine-tune the diffusion model. In the CheXpert classification task, this approach yielded marginal improvements. For example, at 1\% ground truth data, fine-tuned synthetic data improved the F1 score from 0.209 (real-only) to 0.290. However, using only the filtered high-quality synthetic images resulted in lower performance (F1 score = 0.258), suggesting that while filtering improves data quality, the volume and diversity of synthetic data remain critical for downstream performance.

\subsection{Aligning the diffusion model on radiologist feedback yields inconsistent results}
To refine the synthetic image generation process, radiologist feedback was incorporated through a preference-ranking protocol. We found that the fine-tuned synthetic data showed minimal improvements in downstream tasks. For example, in the RSNA Pneumonia classification task, synthetic data fine-tuned with radiologist feedback improved the F1 score from 0.110 (real-only) to 0.118, while the original synthetic data (without fine-tuning) achieved a higher score of 0.146. These results suggest that while radiologist feedback can enhance image quality, its impact on task performance may be limited, particularly when compared to the original synthetic data pipeline.

\section{Conclusion}
This study demonstrates that integrating synthetic chest X-ray images generated via a latent diffusion model can enhance downstream segmentation and classification performance, especially in data-scarce settings. By incorporating paired image-label generation and using feedback from radiologists and proxy models, the framework offers promise in providing a scalable solution for augmenting limited medical datasets. The results confirm that strategically adding synthetic data improves model performance and highlights its potential to address challenges related to dataset scarcity in medical imaging. Moreover, synthetic data can help improve long-tail disease classification by generating additional samples for rare conditions and enhancing the ability of the model to recognize underrepresented diseases.

This study validates that:
\begin{enumerate}
    \item Synthetic data augmentation can enhance performance in both segmentation and classification tasks, particularly in settings with limited ground truth data.
    \item Increasing the volume of synthetic data further correlates positively with improved model performance, demonstrating that larger synthetic datasets can mitigate the challenges of data scarcity.
    \item Generating visually realistic, low-hallucination synthetic images guided by a proxy model can contribute to improved outcomes in downstream tasks.
\end{enumerate}
However, we found that expert feedback is minimally effective in refining model outputs due to the limited size of the expert cohort providing preferences.

Despite these advancements, the study has some limitations. Our evaluation focused exclusively on chest X-rays, which limits understanding of how well the approach would work for other imaging modalities. The tasks we explored were also narrow, as segmentation experiments only targeted pneumothorax. Additionally, relying on a small group of radiologists to collect preference data for fine-tuning may have constrained the model’s optimization, reducing its generalizability. Expanding this work to include diverse imaging types, a broader range of tasks, and feedback from more radiologists could strengthen the effectiveness and applicability of this approach.

\small
\bibliographystyle{ieeenat_fullname}
\bibliography{main}

\end{document}